\newcommand{\be}{\begin{equation}}
\newcommand{\ee}{\end{equation}}
\newcommand{\bfi}{\begin{figure}}
\newcommand{\efi}{\end{figure}}
\newcommand{\bea}{\begin{eqnarray}}
\newcommand{\eea}{\end{eqnarray}}
\newcommand{\beann}{\begin{eqnarray*}}
\newcommand{\eeann}{\end{eqnarray*}}
\newcommand{\beasn}{\begin{sneqnarray}}
\newcommand{\eeasn}{\end{sneqnarray}}
\newcommand{\ba}{\begin{array}}
\newcommand{\ea}{\end{array}}

\newcommand{\Appendix}[1]%
    {\renewcommand{\thesection}{Appendix~\Alph{section}:}%
     \section{#1}}%

\catcode`@=11
\long\def\@makecaption#1#2{
   \vskip 10pt
   \setbox\@tempboxa\hbox{{\small\bf #1.} \ {\small #2}}
   \ifdim \wd\@tempboxa >\hsize       
   {\small\bf #1.} \ {\small #2}\par  
   \else                              
        \hbox to\hsize{\hfil\box\@tempboxa\hfil}
   \fi}
\catcode`@=12

\catcode`@=11
\def\secteqno{\@addtoreset{equation}{section}%
\def\theequation{\thesection.\arabic{equation}}}
\def\endsecteqno{\def\theequation{\@ifundefined{chapter}%
{\arabic{equation}}{\thechapter.\arabic{equation}}}}
\newcounter{subequation}
\def\thesubequation{\alph{subequation}}
\def\sneqnarray{\stepcounter{equation}\let\@currentlabel=\theequation
\setcounter{subequation}{1}
\def\@eqnnum{{\rm (\theequation\thesubequation)}}
\global\@eqcnt\z@\tabskip\@centering\let\\=\@eqncr\let\@@eqncr=\@@sneqncr
$$\halign to \displaywidth\bgroup\@eqnsel\hskip\@centering
 $\displaystyle\tabskip\z@{##}$&\global\@eqcnt\@ne
 \hskip 2\arraycolsep \hfil${##}$\hfil
 &\global\@eqcnt\tw@ \hskip 2\arraycolsep
$\displaystyle\tabskip\z@{##}$\hfil
  \tabskip\@centering&\llap{##}\tabskip\z@\cr}
\def\endsneqnarray{\@@sneqncr\egroup $$\global\@ignoretrue}
\def\@@sneqncr{\let\@tempa\relax
   \ifcase\@eqcnt \def\@tempa{& & &}\or \def\@tempa{& &}
   \else \def\@tempa{&}\fi
     \@tempa \if@eqnsw\@eqnnum\stepcounter{subequation}\fi
     \global\@eqnswtrue\global\@eqcnt\z@\cr}
\def\nobiblabels{\def\@lbibitem[##1]##2{\@bibitem{##2}}}
\catcode`@=12

\def\a{\alpha}  \def\b{\beta}  
\def\d{\delta} \def\D{\Delta} \def\e{\epsilon}
\def\z{\zeta}   
   \def\m{\mu} 
  \def\p{\pi}  
\def\s{\sigma} \def\t{\tau}

\documentclass[12pt]{article}
\usepackage[german,english]{babel}
\usepackage{amssymb}
\usepackage{amsmath}   
\usepackage{braket}
\usepackage{slashed}
\usepackage{multirow}
\usepackage{epsfig}

\secteqno
\renewcommand{\thesection}{\arabic{section}.}

\renewcommand{\theequation}{\thesection  \arabic{equation}}

\begin{document}


\title{{\bf On the quark mass dependence of nucleon--nucleon S--wave scattering lengths}} 
\author{{\Large {\sl Joan Soto}}  {\sl and} {\Large {\sl Jaume Tarr\'us}}\\
        \small{\it{Departament d'Estructura i Constituents de la Mat\`eria 
                   and Institut de Ci\`encies del Cosmos}}\\
        \small{\it{Universitat de Barcelona}}\\
        \small{\it{Diagonal, 647, E-08028 Barcelona, Catalonia, Spain.}}\\  \\
        {\it e-mails:} \small{joan.soto@ub.edu, tarrus@ecm.ub.es}}
\date{\today}

\maketitle

\thispagestyle{empty}

\begin{abstract}
In the framework of a Chiral effective theory with dibaryon fields, we calculate the pion mass dependence of the inverse scattering length of the nucleon--nucleon system in the $^3S_1$ channel at order ${(m_\pi^3 / \Lambda_\chi^2)}\times (m_\pi^{1/2}m_N^{3/2}/8\pi f_\pi^2)^n$ for all $n\ge 0$. 
We show that certain sets of potentially large higher order contributions vanish. We discuss the difficulties of extending the proof to the $^1S_0$ channel. We apply our results to chiral extrapolations of current lattice data.   
\end{abstract}

\section{Introduction}
\indent
The size of the nucleon--nucleon S--wave scattering lengths is larger than expected from standard arguments of chiral counting \cite{Weinberg:1990rz,Weinberg:1991um}, and understanding their values from QCD is still a major challenge \cite{Beane:2011iw}.
Lattice calculations at physical light quark masses are very costly (although they have recently been carried out for some observables \cite{Aoki:2009ix,Durr:2010vn}) and, hence, the use of chiral extrapolations will be needed for some time in order to obtain reliable estimates \cite{Epelbaum:2002gb,Beane:2002xf}. 

The quark mass dependence of low energy observables can be obtained from suitable chiral effective theories. The chiral effective theory for the nucleon--nucleon system was proposed by Weinberg in Refs. \cite{Weinberg:1990rz,Weinberg:1991um}. The fact that the S--wave scattering lengths are unnaturally large, together with other problems of the original proposal related to renormalization and consistency with the chiral counting \cite{Luke:1996hj}, led to the so called KSW (Kaplan, Savage, and Wise) approach \cite{Kaplan:1998tg,Kaplan:1998we} (see \cite{Eiras:2001hu,Beane:2001bc,PavonValderrama:2005gu,Nogga:2005hy,PavonValderrama:2005uj,Yang:2009kx,Yang:2009pn,Beane:2008bt} for alternative approaches, and \cite{Epelbaum:2008ga} for a review on the original approach). It was soon realized that the introduction of dibaryon fields in the effective theory was a very convenient way of implementing large scattering lengths \cite{Kaplan:1996nv,Beane:2000fi}. Nowadays, a nucleon--nucleon effective field theory (NNEFT) with dibaryon fields has been used to calculate the phase shifts in the $^1S_0$ and $^3S_1$--$^3D_1$ channels up to next--to--next--to--leading order \cite{Soto:2007pg,Soto:2009xy}, providing results similar to the KSW approach \cite{Fleming:1999bs,Fleming:1999ee}, with more economical expressions. It was already noticed in \cite{Fleming:1999bs} that beyond next-to--leading order (NLO), part of the calculation must be organized in powers of $\sqrt{m_{\p}/\Lambda_\chi}$, $m_{\p}$ being the pion mass and $\Lambda_\chi$ a typical hadronic scale (say $\Lambda_\chi\sim m_\rho \sim 770$MeV), rather than in powers of $m_{\p}/\Lambda_\chi$ \cite{Soto:2009xy}. It is in fact an accident due to Wigner symmetry that the would--be $\mathcal{O}\left(m^{3/2}_{\p}/\Lambda_{\chi}^{3/2}\right)$ correction vanishes \cite{Mehen:1999qs,Soto:2009xy}. In addition, it was pointed out in Ref. \cite{Mondejar:2006yu} that the terms giving corrections $\sqrt{m_{\p}/\Lambda_\chi}$ were generically large. In this paper we show that these terms can be summed up in the $^3S_1$ channel and, furthermore, that they give a vanishing contribution to the scattering length. This allows us to provide a reliable chiral extrapolation formula for the inverse scattering length including terms up to order $m^{3/2}_q/\Lambda^{1/2}_{\chi}$, $m_q$ being the average light quark masses. Let us recall that the quark mass and the pion mass are related by $m^2_{\p}=2B_0m_q$, where $B_0$ is a low energy constant related to the quark condensate. Unfortunately neither the arguments that allow the resummation nor the proof that the effect vanishes apply to the $^1S_0$ channel. 

The paper is organized as follows. In section $2$ the NNEFT with dibaryon fields is briefly reviewed. In section $3$ we argue that exchanges of potential pions in loops with a radiation pion have to be resummed. We show that they give a vanishing contribution in the $^3S_1$ channel. Section $4$ is devoted to obtaining expressions for the inverse scattering lengths up to $m^{3}_{\p}/\Lambda^2_{\chi}$ terms. In section $5$ we compare our results with the available lattice data. We close with a discussion and conclusions in section $6$.

\section{NNEFT with dibaryon fields}

Our starting point is the effective field theory (EFT) for the $N_B=2$ ($N_B$ being the baryon number) sector of QCD for energies much smaller than $\Lambda_\chi$, proposed in Ref. \cite{Soto:2007pg}. The distinct feature of this EFT is that, in addition to the usual degrees of freedom for a NNEFT theory, namely nucleons and pions, two dibaryon fields, an isovector ($D^a_s$) with quantum numbers $^1 S_0$ and an isoscalar ($\vec{D}_v$) with quantum numbers $^3 S_1$, are also included. Since $m_N \sim \Lambda_\chi$, $m_N$ being the nucleon mass, a non--relativistic formulation of the nucleon fields is convenient \cite{Jenkins:1990jv}. Chiral symmetry, and its breaking due to the quark masses in QCD, constrain the possible interactions of the nucleons and dibaryon fields with the pions. The $N_B=0$ and $N_B=1$ sectors are the usual ones and will be needed only at leading order (LO).

The $N_B=2$ sector consists of terms with (local) two nucleon interactions, dibaryons, and dibaryon--nucleon interactions. The terms with two nucleon interactions can be removed by local field redefinitions \cite{Beane:2000fi} and will not be further considered.  
The LO terms with dibaryon fields and no nucleons in the rest frame of the dibaryons read
\be
\mathcal{L}_{\mathcal{O}(p)}=\frac{1}{2} \text{Tr}\left[D_{s}^{\dag}\Bigl(-id_0+\delta_{m_s}'\Bigr)D_s\right]+
\vec{D}_v^{\dag}\Bigl(-i\partial_0+\delta_{m_v}'\Bigr)\vec{D}_v+ic_{sv}\left(\vec{D}_v^{\dag}\text{Tr}\left[\vec{u} D_s\right]-h.c.\right) \,,
\label{dbLO}
\ee
where $D_s=D^a_s\tau_a$ and $\delta_{m_i}'$, $i=s,v$ are the dibaryon residual masses, which must be much smaller than $\Lambda_\chi$, otherwise the dibaryon should have been integrated out as the remaining resonances were. The covariant derivative for the scalar (isovector) dibaryon field is defined as $d_0D_s=\partial_0D_s+\frac{1}{2}[[u,\partial_0u],D_s]$, 
\be
\begin{split}
\mathcal{L}_{DN}^{(LO)}=& \frac{A_s}{\sqrt{2}}(N^{\dag}\sigma^2\tau^a\tau^2N^*)D_{s,a}+
\frac{A_s}{\sqrt{2}}(N^{\top}\sigma^2\tau^2\tau^aN)D^{\dag}_{s,a}+\\
&+\frac{A_v}{\sqrt{2}}(N^{\dag}\tau^2\vec{\sigma}\sigma^2N^*)\cdot\vec{D}_v+\frac{A_v}{\sqrt{2}}(N^{\top}\tau^2\sigma^2\vec{\sigma} N)\cdot\vec{D}_v^{\dag} \,,
\end{split}
\label{dn}
\ee
with $A_s,A_v\sim \Lambda_\chi^{-1/2}$. The NLO pion--dibaryon
\be
\begin{split}
\mathcal{L}_{\mathcal{O}(p^2)}=
&s_1Tr[D_s(u\mathcal{M}^{\dag}u+u^{\dag}\mathcal{M}u^{\dag})D^{\dag}_s]+s_2Tr[D^{\dag}_s(u\mathcal{M}^{\dag}u+u^{\dag}\mathcal{M}u^{\dag})D_s]+\\
&+v_1\vec{D}^{\dag}_v\cdot\vec{D}_vTr[u^{\dag}\mathcal{M}u^{\dag}+u\mathcal{M}^{\dag}u] +\cdots \,,
\end{split}
\label{orderp2}
\ee
where $\mathcal{M}=m_q \mathbb{I}$. The $s_i$, $i=1,2$, and $v_1$ are low energy constants (LEC). We have only displayed here the terms which will eventually contribute to our calculations. The complete list of operators is given in Appendix B of Ref. \cite{Soto:2009xy}.
The tree level dibaryon propagator expression $i/(-E+\delta_{m_i}'-i\eta)$ gets an important contribution to the self--energy due to the interaction with the nucleons as discussed in Ref. \cite{Soto:2007pg}:
\be
\frac{i}{-E+\delta_{m_i}'+i\frac{A_i^2m_Np}{\pi}}\,, \qquad i=s\,,v\,,
\label{dbself}
\ee
$p=\sqrt{E m_N}$, which is always parametrically larger than the energy $E$. The size of the residual mass can be extracted computing the LO amplitude using the propagator (\ref{dbself}) and matching the result to the effective range expansion,
\be
\delta_{m_i}' \sim \frac{1}{\pi a^i} \sim \frac{m^2_{\p}}{\Lambda_{\chi}}\,, \qquad i=s\,,v\,,
\ee
where $a^i$, $i=s\,, v$, are the scattering lengths of the $^1S_0$ and $^3S_1$ channels respectively. As a consequence $-E+\delta_{m_i}'$ in the full propagator can be expanded for $p\sim m_{\pi}$, and hence the LO expression for the dibaryon field propagator becomes , 
\be
\frac{\pi }{A^2_i m_Np}\,, \qquad i=s\,,v\,.
\label{dbexp}
\ee
The expanded terms can be taken into account through an effective vertex. Moreover, Eq. (\ref{dbself}) implies that the dibaryon field should not be integrated out unless $p\ll \delta_{m_i}'$, instead of $E\ll \delta_{m_i}'$ as the tree level expression suggests. 

In order to calculate the scattering lengths, we need the nucleon--nucleon amplitudes at zero energy. Following  Ref. \cite{Soto:2009xy}, we will first match  NNEFT to pNNEFT, an effective theory for $E\ll m_\pi$ and $p\lesssim m_\pi$, and then match  pNNEFT to $\slashed{\p}$NNEFT, the pionless EFT with dibaryon fields for $p\ll m_\pi$ \cite{Beane:2000fi}, from which we can easily identify the scattering lengths.

\section{Potential pions in loops with radiation pions}

pNNEFT is obtained from NNEFT by integrating out nucleons of energy $E\gtrsim m_\pi$ and pions. Among the latter there are the so called radiation pions, namely pions with $q^0\sim {\bf q}\sim m_\pi$ that interact with nucleons of $E\sim m_\pi$ and $p\sim \sqrt{m_\pi m_N}$. We discuss in this section this particular class of contributions to the matching calculation.

The lowest order diagrams involving radiation pions are depicted in Fig. 6. When a so called potential pion, namely a pion with $q^0\sim m_\pi$ and  ${\bf q}\sim \sqrt{m_\pi m_N}$ in this case, is added to one of those diagrams, for instance as in Fig. 1, a parametric suppression of only $\sqrt{m_{\p}/\Lambda_\chi}$ occurs \cite{Mehen:1999qs,Soto:2009xy}, which numerically turns out to be {\cal O}$(1)$ \cite{Mondejar:2006yu}. It is then necessary to sum up these kinds of contributions. 

\begin{figure}
\centerline{
\includegraphics[height=1.5cm]{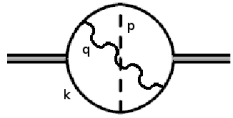}}
\caption{\footnotesize{Example diagram of enhancement of a potential pion inside a radiation pion loop.}}
\label{ex}
\end{figure}

\subsection{Loop resummation}

\begin{figure}
\centerline{\includegraphics[width=14cm]{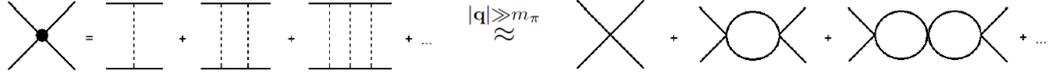}}
\caption{\footnotesize{Potential pion exchanges in the $^1S_0$ channel can be approximated by contact interactions and resummed into an effective vertex when the external momentum is bigger than the pion mass.}}
\label{4neff}
\end{figure}

Let us consider the exchange of \textit{n} potential pions between two nucleon lines. If we project it to the $^1S_0$ channel, the three--momenta coming from the vertices of each potential pion exchange contract between themselves. Note that this is not the case if we project to the $^3S_1$ channel, where a three--momentum from one of the vertices of a given potential pion exchange may get contracted with a three--momentum of a neighboring potential pion exchange vertex. If these \textit{n}--pion exchanges are in a loop with a radiation pion, then the three--momentum in the denominator of the potential pion propagators dominates over the pion mass and the pion energy. As a consequence, the potential pion exchanges collapse into a local vertices (contact interactions) with a coupling constant $g_A^2/(2f^2_{\p})$, where $g_A$ is the axial pion--nucleon coupling constant and $f_{\pi}$ is the pion decay constant as defined in Ref. \cite{Soto:2009xy}. Again, this is not so in the $^3S_1$ channel, where even at very large momentum transfer the potential remains non--local (i.e., it does not reduce to a contact interaction). 
In the left hand side of Fig. \ref{4neff} we depicted the first terms in a series of diagrams with an arbitrary large number of potential pion exchanges. Using the previous reasoning we can collapse the potential pion exchanges into local vertices obtaining the diagrams on the right hand side. In  dimensional regularization the result for the first few terms is
\be
i\frac{g^2_A}{2f^2_{\p}}+i\frac{g^2_A}{2f^2_{\p}}\left(-\frac{\sqrt{q^0-i\e}}{\a}\right)+i\frac{g^2_A}{2f^2_{\p}}\left(-\frac{\sqrt{q^0-i\e}}{\a}\right)^2+\dots\\\,,
\ee
where we have taken the external energy to be $-q^0$, and $\a$ is defined as
\be
\a=\frac{8\p f^2_{\p}}{g_A^2m^{3/2}_N}\,.
\ee
Naively we would expect each bubble to suppress the diagram by a factor of $\sqrt{m_{\p}/\Lambda_{\chi}}$. However a more careful analysis shows that the actual size of each bubble is in fact $\sqrt{m_{\p}}/\a \sim 1.19$, which is of order $\mathcal{O}\left(1\right)$, and hence the series should be resummed. The result of the resummation can be cast as an effective energy--dependent four--nucleon vertex with coupling constant
\be
C_{eff}= i\frac{g^2_A}{2f^2_{\p}}\frac{\a}{\a+\sqrt{q^0-i\e}}\,.
\label{ceff}
\ee

\begin{figure}
\centerline{
\includegraphics[width=14cm]{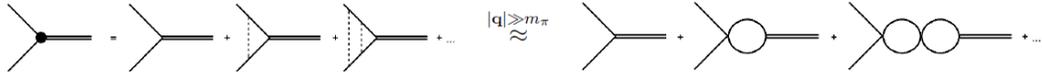}
}
\caption{\footnotesize{Resummation of potential pions in the dibaryon--nucleon vertex.}}
\label{effv}
\end{figure}

An analogous resummation has to be done for potential pion exchanges in the nucleon--dibaryon vertex of Fig. \ref{effv}. Following the same procedure as before, we obtain an energy--dependent effective nucleon--dibaryon vertex,
\be
A_{s,eff}= A_s\frac{\a}{\a+\sqrt{q^0-i\e}}\,.
\ee

\begin{figure}
\centerline{\includegraphics[width=10cm]{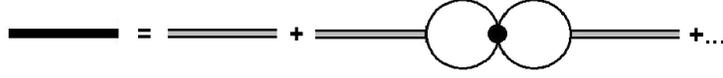}}
\caption{\footnotesize{Inside radiation pion loops the $^1S_0$ receives an additional self--energy contribution.}}
\label{autoenergia}
\end{figure}

Furthermore, using the effective vertex of Eq. (\ref{ceff}) we can construct the self--energy depicted in Fig. \ref{autoenergia}, which inside radiation pion loops turns out to be of order $\mathcal{O}\left(1\right)$ and thus has to be included in the LO propagator (\ref{dbexp}). The following expression for the $^1S_0$ propagator inside radiation pion loops is obtained
\be
-\frac{1}{4A^2_s}\frac{g^2_A}{2f^2_{\p}}\left(1+\frac{\a}{\sqrt{q^0-i\e}}\right)\,.
\ee

Note that in order to have a $^1S_0$ nucleon--nucleon state in a loop with a single radiation pion, the initial nucleon--nucleon state must be in the $^3S_1$ channel. This procedure can then be applied to the calculation of $a^{^3S_1}$, but not to the calculation of $a^{^1S_0}$. This is due to the fact that in the last channel the contact interaction is replaced by a non--local potential that turns out to be singular, and therefore cannot be straightforwardly used in a Lippmann--Schwinger equation; see Refs. \cite{Eiras:2001hu,Beane:2001bc,PavonValderrama:2005gu,Nogga:2005hy,PavonValderrama:2005uj,Yang:2009kx,Yang:2009pn,Beane:2008bt} for discussions and possible solutions.

\subsection[Cancellation of the contributions to a3S1]{Cancellation of the contributions to $a^{^3S_1}$} \label{cancel}

\begin{figure}
\centerline{
\includegraphics[width=10cm]{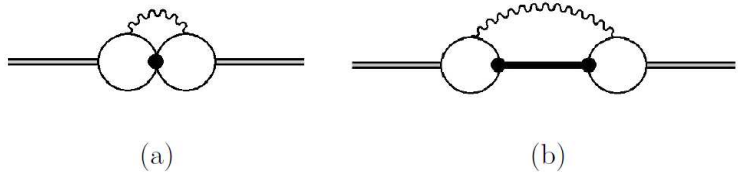}}
\caption{\footnotesize{Order $\mathcal{O}(m^2_\pi/\Lambda_\chi)$ contributions to the dibaryon residual mass.}}
\label{newmatch}
\end{figure}

\begin{figure}
\centerline{\includegraphics[width=10cm]{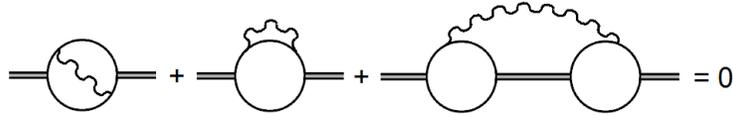}}
\caption{\footnotesize{Order $\mathcal{O}(m^2_\pi/\Lambda_\chi)$ contributions to the dibaryon residual mass with one radiation pion, that cancel due to Wigner symmetry.}}
\label{ws}
\end{figure}

Making use of the new effective vertices obtained by resumming potential pion exchanges, two new diagrams contributing at LO to $a^{^3S_1}$, shown in Fig. \ref{newmatch}, are found:
\be
\mathcal{A}_{a}=8A^2_v \left(\frac{1}{\a}B(1/4,1)-\a^2 B(1,1)+\a B(3/4,1)-B(1/2,1)\right) \,,
\ee
\be
\mathcal{A}_{b}=8A^2_v \Biggl( B(1/2,1)-\a B(3/4,1)+\a^2 B(0,2)-\a^3B(1/4,2)-\a^6B(1,2)+\a^7 B(5/4,2)\Biggr) \,.
\label{d2}
\ee
The definition of $B(\b_1,\b_2)$ can be found in the Appendix. These contributions are of the same order as the diagrams in Fig. \ref{ws}. Those diagrams were counted as $\mathcal{O}\left(m^{5/2}_{\p}/\Lambda^{3/2}_{\chi}\right)$ in Ref. \cite{Soto:2009xy}; however they are proportional to $\sqrt{m_\pi}/\alpha\sim 1$. Then analogous to what we did previously, we should count diagrams in Fig. \ref{ws} as $\mathcal{O}\left(m^{2}_{\p}/\Lambda_{\chi}\right)$. The sum of these diagrams is known to cancel due to Wigner symmetry; however since the third one is already included in Fig. \ref{newmatch}b we should add the first two to Eq. (\ref{d2}) in order to get the complete result at $\mathcal{O}\left(m^{2}_{\p}/\Lambda_{\chi}\right)$,
\be
\mathcal{A}_{s}=-8A^2_v\frac{1}{\a}B(5/4,0) \,.
\ee
The sum of these three contributions ($\mathcal{A}_{a}$, $\mathcal{A}_{b}$, $\mathcal{A}_{s}$) adds up to zero, which can be checked by making use of the relation
\be
B\left(\b_1-1,\b_2\right)=B\left(\b_1,\b_2-1\right)+\a^4 B\left(\b_1,\b_2\right)\,.
\ee
This is at first sight a surprising result. The interaction of nucleons with potential pions spoils the arguments that led to the proof that the sum of the diagrams in Fig. \ref{ws} vanishes as a consequence of Wigner symmetry \cite{Mehen:1999qs}. Yet, since the contact four--nucleon interaction we obtain is only used in the $^1S_0$ channel, it could well be replaced by a Wigner symmetric one with no effect in our calculation, and hence the arguments of Ref. \cite{Mehen:1999qs} would still apply. Nevertheless, as it will become clear soon, the actual reason for the cancellation is that the contact four--nucleon interaction can be removed by the following local field redefinition of the dibaryon field: 
\be
D^a_s \rightarrow D^a_s-\frac{g_A^2}{2f^2_{\p}A_s}N^{T}P_a^{^1S_0}N\,,
\label{fr}
\ee
where $P^{^1S_0}_a=\frac{(i\s_2)(i\t_2\t_a)}{2\sqrt{2}}$, is the projector to the $^1S_0$ partial wave.
Indeed, we have checked that the resummation of potential pion exchanges in the diagrams of Fig. 7, in which Wigner symmetry is violated by the cross and bullet vertices, also vanishes.

As we have mentioned in the previous section, the resummation cannot be carried out for the analogous diagrams for $a^{^1S_0}$. However, it is likely that the perturbative expansion also breaks down in this channel due to numerical factors coming from loop integrals. Hence, any prediction for the quark mass dependence of $a^{^1S_0}$ in terms of a perturbative expansion has to be taken with caution, because it could be missing large corrections. 

Part of the reasoning we have used in the $^3S_1$ channel can be adapted to discuss the result for the diagrams with a single potential pion exchange in a loop with a radiation pion in the $^1S_0$ channel. In this set of diagrams, the radiation pion three--momentum in the denominators of the loop integral can be neglected in front of any of the nucleons or potential pion three--momenta, so the potential pion three momenta in the pion--nucleon vertices must end up contracted between themselves, and hence we are left with a situation analogous to the one in the $^3S_1$ channel. At this point we can approximate the potential exchange by a four--nucleon contact term, following the same reasoning as for the contributions to $a^{^3S_1}$. The contact term can then be eliminated by a field redefinition analogous to Eq. (\ref{fr}) for the ${\vec D}_v$ dibaryon field. We then conclude that the sum of the diagrams in Fig. \ref{ws} with a single potential pion insertion must also vanish in the $^1S_0$ channel. This result is in contradiction with those of Refs. \cite{Fleming:1999ee,Soto:2009xy}, where this class of diagrams with one potential pion inside a radiation pion loop was found to be non--zero. We believe that this is  a consequence of double counting certain diagrams. In particular, the last diagram in Fig. 17 of Ref. \cite{Fleming:1999ee} is already included in the first one. According to our calculations this error would lead to the result presented in Ref. \cite{Fleming:1999ee,Soto:2009xy}.

\section{Scattering Lengths}

In this section we sketch the matching between pNNEFT and NNEFT, and between pNNEFT and $\slashed{\p}$NNEFT in the light of results of the previous section. In particular we focus our efforts on obtaining NLO expressions for the residual mass and the dibaryon--nucleon vertices' low energy constants. With these expressions we write the scattering lengths up to $m^{3}_{\p}/\Lambda^2_{\chi}$ terms. Results for the $^1S_0$ channel have to be taken with caution due to possible large corrections from multiple potential pion exchanges in the loops with a radiation pion, as explained in the previous section. The following subsections are rather sketchy. We refer the reader to section 4 of Ref. \cite{Soto:2009xy} for details on the matching procedure beween NNEFT and pNNEFT, and to section 6 of the same reference for details on the one between pNNEFT and $\slashed{\p}$NNEFT.

\subsection{Matching pNNEFT with NNEFT}

\begin{figure}
\centerline{\includegraphics[width=10cm]{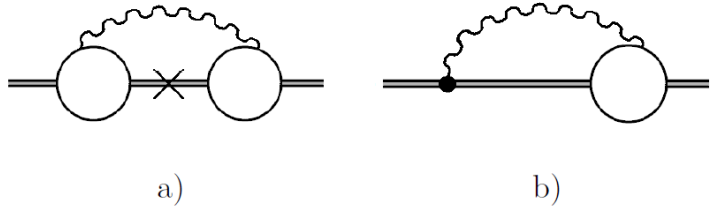}
}
\caption{\footnotesize{Order $\mathcal{O}(m^3_\pi/\Lambda^2_\chi)$ contributions to the dibaryon residual mass.}}
\label{matching1}
\end{figure}

In the one--nucleon sector, pion loops produce a shift in the nucleon mass, $\d m_N$, that introduces a quark mass dependence. We can reshuffle $\d m_N$ into the dibaryon residual mass by local field redefinitions. The expression for $\d m_N$ can be found \cite{Procura:2003ig}, and up to $\mathcal{O}\left(m^3_{\p}/\Lambda^2_{\chi}\right)$ contributions it reads
\be
\d m_N=-4c_1 m^2_{\p}-\frac{3g^2_A}{32\p f^2_{\p}}m^3_{\p} \,.
\ee
In the NLO pion--dibaryon Lagrangian (\ref{orderp2}) the residual mass gets $\mathcal{O}\left(m^2_{\p}/\Lambda_{\chi}\right)$ contributions proportional to the quark mass. Additional $\mathcal{O}\left(m^3_{\p}/\Lambda^2_{\chi}\right)$ contributions come from the diagrams in Fig. \ref{matching1}. Adding up all the contributions we obtain the formula for the residual mass,
\be
\begin{split}
\d_{m_v}&=\d_{m_v}'+2\frac{v_1}{B_0}m^2_{\p}+2\d m_N+\left(\frac{g^2_A}{2f^2_{\p}}\right)\frac{m^3_{\p}}{8\p}\frac{A^2_v}{A^2_s}+c_{sv}\left(\frac{g_A}{f^2_{\p}}\right)\frac{m^3_{\p}}{8\p}\frac{A_v}{A_s}\,, \\
\d_{m_s}&=\d_{m_s}'+2\frac{s_1+s_2}{B_0}m^2_{\p}+2\d m_N+\left(\frac{g^2_A}{2f^2_{\p}}\right)\frac{m^3_{\p}}{8\p}\frac{A^2_s}{A^2_v}+c_{sv}\left(\frac{g_A}{f^2_{\p}}\right)\frac{m^3_{\p}}{8\p}\frac{A_s}{A_v}\,. \\
\end{split}
\label{pn}
\ee
In the two--nucleon sector only the one--pion exchange is relevant at this order, which produces the well known one--pion exchange potential.

\subsection[Matching pNNEFT with pionless NNEFT]{Matching pNNEFT with $\slashed{\p}$NNEFT}

\begin{figure}
\centerline{\includegraphics[width=10cm]{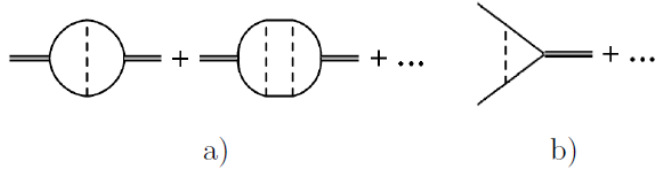}
}
\caption{\footnotesize{Contributions to the matching between pNNEFT and $\slashed{\p}$NNEFT. (a) Leading and NLO contributions to the residual mass. (b) NLO correction to the dibaryon--nucleon vertex low--energy constant. Dashed lines represent the one--pion--exchange potential.}}
\label{matching2}
\end{figure}

The next step in order to evaluate the scattering length is to build a theory valid for $p\lesssim\d_m$. This is achieved by integrating out the nucleon three--momenta of order $m_{\p}$, which leads to the so--called pionless nucleon--nucleon EFT. Non--local potentials can be expanded in powers of $\frac{p^2}{m^2_{\p}}$ and become local. Self energies in Fig. \ref{matching2}(a) can be expanded, giving contributions to the dibaryon residual mass as well as time--derivative terms. The latter can be reabsorbed by field redefinitions of the dibaryon fields. The dibaryon--nucleon vertex gets contributions from the diagrams in Fig. \ref{matching2}(b). Recall that the one pion exchange potentials in Fig. \ref{matching2} correspond to potential pions with ${\bf q}\sim m_\pi$ and not to potential pions with ${\bf q}\sim \sqrt{m_\pi m_N}$, such as the ones considered in section 3.

The contribution to the residual mass from the first diagram in Fig. \ref{matching2}(a) is of order $\mathcal{O}\left(m^{2}_{\p}/\Lambda_{\chi}\right)$. This diagram contains a divergence proportional to the quark mass which is renormalized by the counterterm of the same order proportional to the quark mass in Eq. (\ref{pn}). The second diagram in Fig. \ref{matching2}(a) is NLO, $\mathcal{O}\left(m^{3}_{\p}/\Lambda^2_{\chi}\right)$:
 
\be
\begin{split}
\d^{LO}_{m_v}&=\d_{m_v}'+2 \frac{v_1}{B_0} m^2_{\p}-8c_1 m^2_{\p}-A^2_v\frac{g^2_A}{f^2_{\p}}\left(\frac{m_{\p}m_N}{4\p}\right)^2\ln\left(\frac{m^2_{\p}}{\m^2}\right)\,, \\
\d^{LO}_{m_s}&=\d_{m_s}'+2\frac{s_1+s_2}{B_0}m^2_{\p}-8c_1 m^2_{\p}-A^2_s\frac{g^2_A}{f^2_{\p}}\left(\frac{m_{\p}m_N}{4\p}\right)^2\ln\left(\frac{m^2_{\p}}{\m^2}\right)\,, \\
\end{split}
\ee
\be
\begin{split}
\d^{NLO}_{m_v}&=-\frac{3g^2_A}{16\p f^2_{\p}}m^3_{\p}+\left(\frac{g^2_A}{2f^2_{\p}}\right)\frac{m^3_{\p}}{8\p}\frac{A^2_v}{A^2_s}+c_{sv}\left(\frac{g_A}{f^2_{\p}}\right)\frac{m^3_{\p}}{8\p}\frac{A_v}{A_s}+A^2_v\left(\frac{g^2_A}{f^2_{\p}}\right)^2\left(\frac{m_{\p}m_N}{4\p}\right)^3\frac{5}{2}\left(6+13\ln\left(2\right)\right)\,, \\
\d^{NLO}_{m_s}&=-\frac{3g^2_A}{16\p f^2_{\p}}m^3_{\p}+\left(\frac{g^2_A}{2f^2_{\p}}\right)\frac{m^3_{\p}}{8\p}\frac{A^2_s}{A^2_v}+c_{sv}\left(\frac{g_A}{f^2_{\p}}\right)\frac{m^3_{\p}}{8\p}\frac{A_s}{A_v}+A^2_s\left(\frac{g^2_A}{f^2_{\p}}\right)^2\left(\frac{m_{\p}m_N}{4\p}\right)^32\ln\left(2\right)\,. \\
\end{split}
\ee
The dibaryon--nucleon vertex up to NLO gets only one new contribution from the first diagram in Fig. \ref{matching2}(b). Defining $A^2_{i,NLO}=A^2_i\D_{NLO}\,\,,i=s,v$,
\be
\D_{NLO}=\frac{g^2_A}{2f_{\p}^2}\frac{m_{\p}m_N}{4\p}\,.
\ee
Note that the parametric suppression of an extra pion exchange in the diagrams of Fig. \ref{matching2} is $m_\pi m_N/\Lambda_\chi^2$ whereas the one in the diagrams in Figs. \ref{4neff} and \ref{effv} is $m_\pi^{1/2} m_N^{3/2}/\Lambda_\chi^2$. Hence the resummation of diagrams in Fig. \ref{matching2} is less important than the ones in Figs. \ref{4neff} and \ref{effv}, a feature that justifies why in our power counting scheme they need not be resummed.

Now we have all the ingredients to write the expression for the scattering lengths:
\be
a^{-1}_{i}=\frac{\p\d^{LO}_{m_i}}{m_N A^2_i}\left(1-\D_{NLO}\right)+\frac{\p\d^{NLO}_{m_i}}{m_N A^2_i} \quad,i=s(^1S_0)\,,v(^3S_1)\,.
\ee

\section{Comparison with lattice data} \label{fl}

The expressions for the scattering lengths can be rewritten to collect all the parameters into three independent ones, 
\be
\begin{split}
a^{-1}_{i}=&\z_{i1}\left(1-\frac{g^2_A m_N}{8 \p f^2_{\p}}m_{\p}\right)+\left[\z_{i2}-\frac{g_A^2 m_N}{16\p f^2_{\p}}\ln\left(\frac{m^2_{\p}}{\m^2}\right)\right]m^2_{\p}+\z_{i3}m^3_{\p} \\
&+\frac{1}{2}\left(\frac{g^2_A m_N}{8\p f^2_{\p}}\right)^2 m^3_{\p}\ln\left(\frac{m^2_{\p}}{\m^2}\right) \quad,\quad i=s(^1S_0)\,,v(^3S_1)\,.
\end{split}
\label{qex}
\ee
The expression obtained is quite simple and emphasizes the $m_{\p}$ dependence. The relation of the $\z$ parameters to the low energy constants of the EFT can be found in Table \ref{t1}. The expected sizes of these parameter are, $\z_{i1}\sim \mathcal{O}\left(m^2_{\p}/\Lambda_{\chi}\right)$, $\z_{i2}\sim \mathcal{O}\left(1/\Lambda_{\chi}\right)$ and $\z_{i3}\sim \mathcal{O}\left(1/\Lambda^2_{\chi}\right)$.
\begin{table}
\centerline{
\begin{tabular}{|c|c|c|c|} \hline\hline
        & $\z_1$ & $\z_2$ & $\z_3$ \\ \hline
$^1S_0$ & $\frac{\p \d_{m_s}'}{m_N A^2_s}$ & $\frac{2\p((s_1+s_2)/B_0-8c_1)}{m_N A^2_s}$ &  $\frac{g_A^2}{16 m_N f^2_{\p}}\left(\frac{1}{A^2_v}+\frac{2c_{sv}}{g_A A_s A_v}-\frac{3}{A^2_s}\right)-\frac{g_A^2}{4f^2_{\p}}\frac{(s_1+s_2)/B_0-8c_1}{ A^2_s}+\left(\frac{g^2_A m_N}{f^2_{\p}}\right)^2\frac{\log(2)}{128\p^2}$\\ 
$^3S_1$ & $\frac{\p \d_{m_v}'}{m_N A^2_v}$ & $\frac{2\p(v_1/B_0-8c_1)}{m_N A^2_s}$ & $\frac{g_A^2}{16 m_N f^2_{\p}}\left(\frac{1}{A^2_s}+\frac{2c_{sv}}{g_A A_s A_v}-\frac{3}{A^2_v}\right)-\frac{g_A^2}{4f^2_{\p}}\frac{(v_1/B_0-8c_1)}{A^2_v}+5\left(\frac{g^2_A m_N}{f^2_{\p}}\right)^2\frac{6+13\log(2)}{256\p^2}$      \\  \hline\hline
\end{tabular}}
\caption{\footnotesize{Independent free parameters in terms of the effective theory low energy constants.}}
\label{t1}
\end{table}

\begin{figure}
\centerline{\begin{tabular}{cc}
\includegraphics[width=7cm]{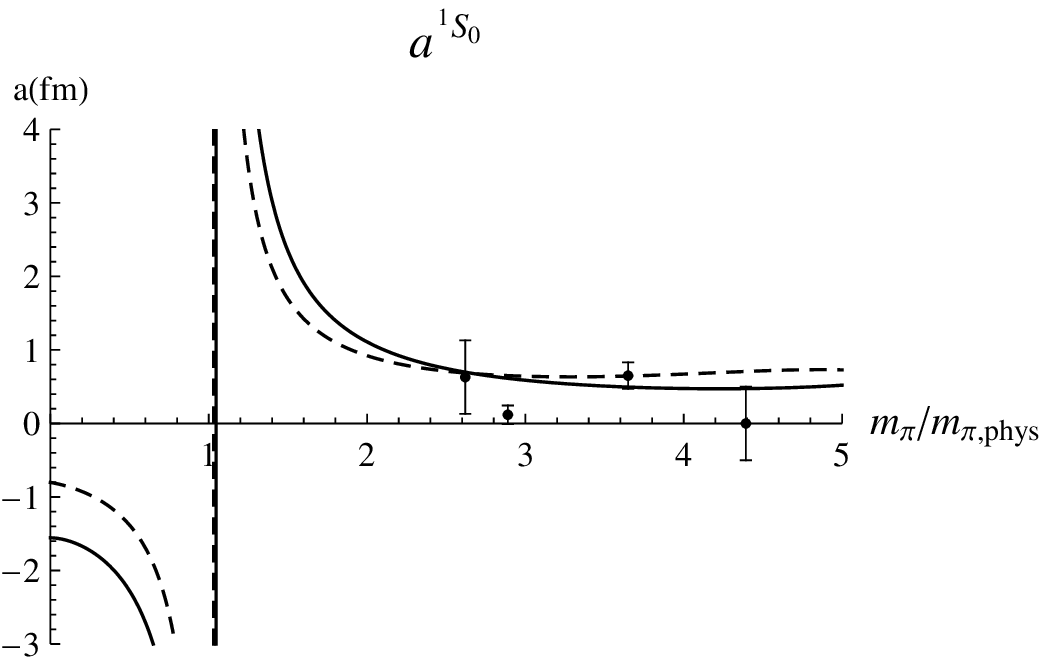} & \includegraphics[width=7cm]{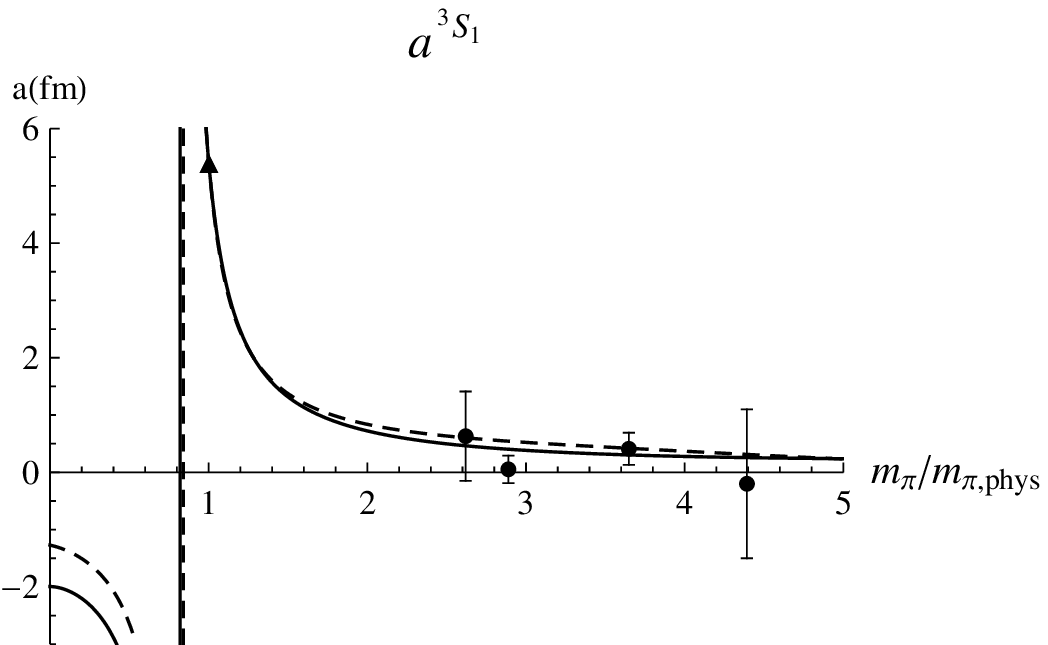} \\
\end{tabular}}
\caption{\footnotesize{Plots of $a^{^1S_0}$ (left) and $a^{^3S_1}$ (right). The solid and dashed lines correspond to the LO and NLO respectively. The triangular dot in the $a^{^3S_1}$ figure corresponds the physical value of the scattering length. In the $a^{^1S_0}$ figure the physical point is out of scale.}}
\label{plots}
\end{figure}

The first lattice QCD calculation of the nucleon--nucleon scattering lengths was performed by Fukugita \textit{et al} \cite{Fukugita:1994na,Fukugita:1994ve} in the quenched approximation with Wilson quark action. More recent studies using the quenched approximation have been carried out by Aoki \textit{et al} \cite{Aoki:2008hh}. The NPLQCD Collaboration has performed unquenched calculations in mixed--action (domain wall--staggered) \cite{Beane:2006mx} and anisotropic clover--quark action \cite{Beane:2009py}. 
\begin{table}
\centerline{\begin{tabular}{|c|c|c|}\hline\hline
 $m_{\p}(MeV)$ & $a^{^1S_0}(fm)$ & $a^{^3S_1}(fm)$ \\ \hline
 $353.7$ & $0.63 \pm 0.50$ & $0.63 \pm 0.74$  \\
 $492.5$ & $0.65 \pm 0.18$ & $0.41 \pm 0.28$ \\
 $593$   & $0.0  \pm 0.5$  & $-0.2 \pm 1.3$\\
 $390$   & $0.118^{+0.109}_{-0.126}$ & $0.052^{+0.18}_{-0.24}$ \\ \hline\hline
\end{tabular}}
\caption{\footnotesize{Lattice data point used to fit the scattering lengths. The first three data points are from Ref. \cite{Beane:2006mx} and the fourth one is from Ref. \cite{Beane:2009py}.}}
\label{data}
\end{table}
We fitted the lattice data of the NPLQCD Collaboration (see Table \ref{data}). Unfortunately all data points are above or close to $350\,MeV$, a scale beyond which it is not clear that chiral extrapolations for the nucleon--nucleon system are still valid. Thus the obtained results have to be taken with caution. We forced the expressions for the scattering lengths to reproduce the experimental values at the physical pion mass, $a^{^1S_0}=-23.7$ fm and $a^{^3S_1}=5.38$ fm. This allowed us to solve one parameter as a function the remaining ones, we chose to solve $\z_{i1}$. The remaining parameters have been obtained by minimizing an augmented chi--square distribution \cite{Schindler:2008fh} for each scattering length. The augmented chi--square distribution is defined as the sum of the chi--square function with a set of priors for every one of the free parameters to be fitted,
\be
\begin{split}
&\chi^2_{aug}=\chi^2_{a^i}+\chi^2_{prior}\,, \\
\chi^2_{a^i}=\frac{1}{n}\sum\limits^n_{j=1} \frac{\left(a^i(m_{\p,j})-a^i_j\right)^2 }{\d^2_{a^i_j}} \quad ,& \quad \chi^2_{prior}=\frac{1}{N}\sum\limits^N_{k=1} \frac{\left(\ln \|x_k\| -\ln \|\bar{x}_k\|\right)^2 }{\ln^2 R_k} \quad,\quad i= ^1S_0, ^3S_1\,, 
\end{split}
\ee
where $a^i_j$ and $\d_{a^i_j}$ stand for the value of scattering length and its uncertainity  at the pion mass $m_{\p,j}$ respectively. $n$ is the total number of lattice data points. Furthermore, $x_k$ refers to the free parameters, $N$ being their total number. The free parameters are $\z_{i2}$ at LO, and $\z_{i2}$ and $\z_{i3}$ at NLO. The prior information is obtained from naive dimensional analysis. For instance, if the parameter $x_k$ is of order ${\cal O}(1)$, we would expect it to be in the range $0.1<\|x_k\|<10$, which translates to setting $\ln(\|\bar{x}_k\|)=0$ and $\ln(R_k)=1$ for the \textit{k}th parameter. We have taken logarithms in the prior functions to achieve equal weights for the subranges $0.1<\|x_k\|<1$ and $1<\|x_k\|<10$. For $\z_{i2}$, priors are set to $\bar{\z}_{i2}=\frac{1}{\Lambda_\chi}$ and $\ln\left(R_{\z_{i2}}\right)=1$, and for $\z_{i3}$ to $\bar{\z}_{i3}=\frac{1}{\Lambda^2_\chi}$ and $\ln\left(R_{\z_{i3}}\right)=1$. The plots corresponding to the fits of the leading and next--to--leading order expressions of the scattering lengths as a function of the quark mass are displayed in Fig. \ref{plots}. The chi--squared distribution per degree of freedom is defined as
\be
\chi^2_{a^i,d.o.f}=\frac{1}{n-1-N}\sum\limits^n_{j=1} \frac{\left(a^i(m_{\p,j})-a^i_j\right)^2 }{\d^2_{a^i_j}} \quad,\quad i= ^1S_0, ^3S_1\,.
\ee
The values obtained for the parameters and the chi--squared per degree of freedom are collected in Tables \ref{paralo} and \ref{paranlo}. The values obtained for $\z_{s1}$ and $\z_{i3}$, $i=s,v$, at NLO are on the limit of what we would consider natural size. This could indicate that significant cancellations occur at the physical pion mass in order to produce the observed values of the scattering lengths. Note that the fine tuning increases with the precision of the expression used.
 
\begin{table}
\centerline{\begin{tabular}{|c|c|c|c|}\hline\hline
    LO   & $\chi_{d.o.f}^2$ & $\z_1(MeV)$ & $\z_2(MeV^{-1})$   \\ \hline
 $^1S_0$ & $3.74$   & $-126$      & $0.67\cdot10^{-3}$ \\
 $^3S_1$ & $0.91$   & $-98$       & $1.59\cdot10^{-3}$ \\ \hline\hline
\end{tabular}}
\caption{\footnotesize{LO fit parameters results.}}
\label{paralo}
\end{table}

\begin{table}
\centerline{\begin{tabular}{|c|c|c|c|c|}\hline\hline
   NLO  & $\chi_{d.o.f}^2$ & $\z_1(MeV)$ & $\z_2(MeV^{-1})$   & $\z_3(MeV^{-2})$   \\ \hline
 $^1S_0$& $2.4$   & $-246$      & $4.56\cdot10^{-3}$ & $9.21\cdot10^{-6}$ \\
 $^3S_1$& $0.4 $  & $-155$      & $3.83\cdot10^{-3}$ & $10.1\cdot10^{-6}$ \\ \hline\hline
\end{tabular}}
\caption{\footnotesize{NLO fit parameters results.}}
\label{paranlo}
\end{table}

The $m_q$--dependence of the scattering lengths has been studied previously in Refs. \cite{Epelbaum:2002gb,Epelbaum:2002gk} using numerical solutions to the Lippmann--Schwinger equation with potentials obtained from Weinbergs's power counting, and in Refs. \cite{Beane:2002xf,Beane:2001bc} in the framework of BBSvK counting. All these papers were written before the first unquenched lattice results appeared and hence do not use lattice data to fit their unknown free parameters. In both approaches the behavior of the scattering length was studied for a suitable range of the unknown parameters. Special attention was devoted to the extrapolations to the chiral limit. A more recent study can be found in Ref. \cite{Chen:2010yt} using the power counting of Ref. \cite{Beane:2008bt} and lattice data of the NPLQCD Collaboration. In the $^1S_0$ channel our results in the chiral limit indicate that the scattering length remains negative, thus the system is unbounded, coinciding with the predictions of mentioned previous works, albeit our value seems slightly smaller. In the $^3S_1$ channel our extrapolation of the scattering length to the chiral limit shows that it evolves from positive values at the physical pion mass to negative values, hence going from a bounded nucleon--nucleon system to an unbounded one. This is opposite to the results in Refs. \cite{Epelbaum:2002gb,Epelbaum:2002gk}, and to those in Refs. \cite{Beane:2002xf,Beane:2001bc}, for most of the parameter space, in which the scattering length remains positive in the whole range from the chiral limit to the physical pion mass. Nevertheless, in Refs. \cite{Beane:2002xf,Beane:2001bc} a behavior similar to the one we have obtained is observed in certain regions of the parameter space. In Ref. \cite{Chen:2010yt}, the only one of the pevious works on the $m_q$--dependence of the scattering lengths that has used lattice data, the $^3S_1$ channel goes to negative values in the chiral limit, and overall presents a very similar result to ours.

\section{Conclusions}

We have showed that certain classes of diagrams involving potential pion exchanges in loops with radiation pions 
can be summed up in the $^3S_1$ channel. This is important because each of these exchanges introduces a parametric suppression of only $\mathcal{ O}(\sqrt{m_\pi / m_N})$ that numerically turns out to be $\mathcal{O}(1)$. The resummation is possible because after radiating a pion a nucleon--nucleon system in the $^3S_1$ channel changes into the $^1S_0$ channel, and in this channel the one--pion exchange potential at high momentum transfer becomes a contact interaction. We showed that by performing dibaryon local field redefinitions we can get rid of the contact interaction, and hence the contribution of all diagrams involving these potential pion exchanges must be zero. We checked this cancellation by explicitly computing the diagrams and adding them up.

Unfortunately, in the $^1S_0$ channel it has not been possible for us to compute the contribution of an arbitrary number of potential pions in a loop with a radiation pion. This is because after radiating a pion a nucleon--nucleon system in the $^1S_0$ channel changes into the $^3S_1$ channel, and in this channel the one--pion exchange potential at high momentum transfer does not reduce to a contact interaction anymore. However, similar arguments still apply to the diagrams with only one potential pion, which should then add up to zero. This is in contradiction with the results of Refs. \cite{Soto:2009xy,Fleming:1999ee}, and we have pointed out a possible source of the discrepancy in section \ref{cancel}. It is very likely that in the $^1S_0$ channel the perturbative series breaks down as in the $^3S_1$ channel, which means that it is possible that our expressions for $a^{^1S_0}$ are missing large contributions, and hence, are unreliable.

We have given chiral extrapolation formulas for $1/a^{^1S_0}$ and $1/a^{^3S_1}$ up to corrections of order $\mathcal{O}\left(m^{3}_{\p}/\Lambda^2_{\chi}\right)$ depending on three independent free parameters. In section \ref{fl} we carried out a fit of these expressions to lattice data from the NPLQCD Collaboration \cite{Beane:2006mx,Beane:2009py}. The results in Fig. \ref{plots}, Table \ref{paralo} and Table \ref{paranlo} show that our expressions for $a^{^3S_1}$ are much more compatible with lattice data than those for $a^{^1S_0}$, which could indicate that the missing, potentially large, contributions to $a^{^1S_0}$ previously mentioned do exist. Using these results to extrapolate the scattering lengths in the chiral limit, we obtain that $a^{^1S_0}$ keeps its negative sign, while $a^{^3S_1}$ changes from positive to negative. However, at this stage, lattice data sets available are rather small, with relatively large pion masses, and often computed using different approaches, making it difficult hold any strong statement in this respect.

\bigskip

{\bf Acknowledgments}

\bigskip

We have been supported by the CPAN  CSD2007-00042 Consolider--Ingenio 2010 program (Spain), the 2009SGR502 CUR grant (Catalonia) and the FPA2010-16963 project (Spain). JT acknowledges a MEC FPU grant (Spain).

\begin{center}
\section*{Appendix}
\end{center}

The $B(\b_1,\b_2)$ loop integrals used in section \ref{cancel} are defined as follows:
\be
B(\b_1,\b_2)=(\mu^2)^{d-4}\int \frac{d^{d-1}q}{(4\p)^{d-1}}\frac{\mathbf{q}^2}{(\mathbf{q}^2+m^2_{\p})^{\b_1}}\frac{1}{(\mathbf{q}^2+m^2_{\p}-\a^4)^{\b_2}}\,.
\ee 


\begin{thebibliography}{99}

\bibitem{Weinberg:1990rz}
  S.~Weinberg,
  Phys.\ Lett.\  B {\bf 251}, 288 (1990).

\bibitem{Weinberg:1991um}
  S.~Weinberg,
  Nucl.\ Phys.\  B {\bf 363}, 3 (1991).

\bibitem{Beane:2011iw}
  S.~R.~Beane {\it et al.}  [NPLQCD Collaboration],
  arXiv:1109.2889 [hep-lat].

\bibitem{Aoki:2009ix}
  S.~Aoki {\it et al.}  [PACS-CS Collaboration],
  Phys.\ Rev.\  D {\bf 81}, 074503 (2010)
  [arXiv:0911.2561 [hep-lat]].

\bibitem{Durr:2010vn}
  S.~Durr {\it et al.},
  Phys.\ Lett.\  B {\bf 701}, 265 (2011)
  [arXiv:1011.2403 [hep-lat]].

\bibitem{Epelbaum:2002gb}
  E.~Epelbaum, U.~G.~Meissner and W.~Gloeckle,
  Nucl.\ Phys.\  A {\bf 714}, 535 (2003)
  [arXiv:nucl-th/0207089].

\bibitem{Beane:2002xf}
  S.~R.~Beane and M.~J.~Savage,
  Nucl.\ Phys.\  A {\bf 717}, 91 (2003)
  [arXiv:nucl-th/0208021].

\bibitem{Luke:1996hj}
  M.~E.~Luke and A.~V.~Manohar,
  Phys.\ Rev.\  D {\bf 55}, 4129 (1997)
  [arXiv:hep-ph/9610534].

\bibitem{Kaplan:1998tg}
  D.~B.~Kaplan, M.~J.~Savage and M.~B.~Wise,
  Phys.\ Lett.\  B {\bf 424}, 390 (1998)
  [arXiv:nucl-th/9801034].

\bibitem{Kaplan:1998we}
  D.~B.~Kaplan, M.~J.~Savage and M.~B.~Wise,
  Nucl.\ Phys.\  B {\bf 534}, 329 (1998)
  [arXiv:nucl-th/9802075].

\bibitem{Eiras:2001hu}
  D.~Eiras and J.~Soto,
  Eur.\ Phys.\ J.\  A {\bf 17}, 89 (2003)
  [arXiv:nucl-th/0107009].

\bibitem{Beane:2001bc}
  S.~R.~Beane, P.~F.~Bedaque, M.~J.~Savage and U.~van Kolck,
  Nucl.\ Phys.\  A {\bf 700}, 377 (2002)
  [arXiv:nucl-th/0104030].

\bibitem{PavonValderrama:2005gu}
  M.~Pavon Valderrama and E.~Ruiz Arriola,
  Phys.\ Rev.\  C {\bf 72}, 054002 (2005)
  [arXiv:nucl-th/0504067].

\bibitem{Nogga:2005hy}
  A.~Nogga, R.~G.~E.~Timmermans and U.~van Kolck,
  Phys.\ Rev.\  C {\bf 72}, 054006 (2005)
  [arXiv:nucl-th/0506005].

\bibitem{PavonValderrama:2005uj}
  M.~Pavon Valderrama and E.~Ruiz Arriola,
  Phys.\ Rev.\  C {\bf 74}, 064004 (2006)
  [Erratum-ibid.\  C {\bf 75}, 059905 (2007)]
  [arXiv:nucl-th/0507075].

\bibitem{Yang:2009kx}
  C.~J.~Yang, C.~Elster and D.~R.~Phillips,
  Phys.\ Rev.\  C {\bf 80}, 034002 (2009)
  [arXiv:0901.2663 [nucl-th]].

\bibitem{Yang:2009pn}
  C.~J.~Yang, C.~Elster and D.~R.~Phillips,
  Phys.\ Rev.\  C {\bf 80}, 044002 (2009)
  [arXiv:0905.4943 [nucl-th]].

\bibitem{Beane:2008bt}
  S.~R.~Beane, D.~B.~Kaplan and A.~Vuorinen,
  Phys.\ Rev.\  C {\bf 80}, 011001 (2009)
  [arXiv:0812.3938 [nucl-th]].

\bibitem{Epelbaum:2008ga}
  E.~Epelbaum, H.~W.~Hammer and U.~G.~Meissner,
  Rev.\ Mod.\ Phys.\  {\bf 81}, 1773 (2009)
  [arXiv:0811.1338 [nucl-th]].

\bibitem{Kaplan:1996nv}
  D.~B.~Kaplan,
  Nucl.\ Phys.\  B {\bf 494}, 471 (1997)
  [arXiv:nucl-th/9610052].

\bibitem{Beane:2000fi}
  S.~R.~Beane and M.~J.~Savage,
  Nucl.\ Phys.\  A {\bf 694}, 511 (2001)
  [arXiv:nucl-th/0011067].

\bibitem{Soto:2007pg}
  J.~Soto and J.~Tarrus,
  Phys.\ Rev.\  C {\bf 78}, 024003 (2008)
  [arXiv:0712.3404 [nucl-th]].

\bibitem{Soto:2009xy}
  J.~Soto and J.~Tarrus,
  Phys.\ Rev.\  C {\bf 81}, 014005 (2010)
  [arXiv:0906.1194 [nucl-th]].

\bibitem{Fleming:1999bs}
  S.~Fleming, T.~Mehen and I.~W.~Stewart,
  Phys.\ Rev.\  C {\bf 61}, 044005 (2000)
  [arXiv:nucl-th/9906056].

\bibitem{Fleming:1999ee}
  S.~Fleming, T.~Mehen and I.~W.~Stewart,
  Nucl.\ Phys.\  A {\bf 677}, 313 (2000)
  [arXiv:nucl-th/9911001].

\bibitem{Mehen:1999qs}
  T.~Mehen, I.~W.~Stewart and M.~B.~Wise,
  Phys.\ Rev.\ Lett.\  {\bf 83}, 931 (1999)
  [arXiv:hep-ph/9902370].

\bibitem{Mondejar:2006yu}
  J.~Mondejar and J.~Soto,
  Eur.\ Phys.\ J.\  A {\bf 32}, 77 (2007)
  [arXiv:nucl-th/0612051].

\bibitem{Jenkins:1990jv}
  E.~E.~Jenkins and A.~V.~Manohar,
  Phys.\ Lett.\  B {\bf 255}, 558 (1991).

\bibitem{Procura:2003ig}
  M.~Procura, T.~R.~Hemmert and W.~Weise,
  Phys.\ Rev.\  D {\bf 69}, 034505 (2004)
  [arXiv:hep-lat/0309020].

\bibitem{Fukugita:1994na}
  M.~Fukugita, Y.~Kuramashi, H.~Mino, M.~Okawa and A.~Ukawa,
  Phys.\ Rev.\ Lett.\  {\bf 73}, 2176 (1994)
  [arXiv:hep-lat/9407012].

\bibitem{Fukugita:1994ve}
  M.~Fukugita, Y.~Kuramashi, M.~Okawa, H.~Mino and A.~Ukawa,
  Phys.\ Rev.\  D {\bf 52}, 3003 (1995)
  [arXiv:hep-lat/9501024].

\bibitem{Aoki:2008hh}
  S.~Aoki, T.~Hatsuda and N.~Ishii,
  Comput.\ Sci.\ Dis.\  {\bf 1}, 015009 (2008)
  [arXiv:0805.2462 [hep-ph]].

\bibitem{Beane:2006mx}
  S.~R.~Beane, P.~F.~Bedaque, K.~Orginos and M.~J.~Savage,
  Phys.\ Rev.\ Lett.\  {\bf 97}, 012001 (2006)
  [arXiv:hep-lat/0602010].

\bibitem{Beane:2009py}
  S.~R.~Beane {\it et al.}  [NPLQCD Collaboration],
  Phys.\ Rev.\  D {\bf 81}, 054505 (2010)
  [arXiv:0912.4243 [hep-lat]].

\bibitem{Schindler:2008fh}
  M.~R.~Schindler and D.~R.~Phillips,
  Annals Phys.\  {\bf 324}, 682 (2009)
  [Erratum-ibid.\  {\bf 324}, 2051 (2009)]
  [arXiv:0808.3643 [hep-ph]].

\bibitem{Epelbaum:2002gk}
  E.~Epelbaum, U.~G.~Meissner and W.~Gloeckle,
  arXiv:nucl-th/0208040.
  
\bibitem{Chen:2010yt}
  J.~-W.~Chen, T.~-K.~Lee, C.~-P.~Liu and Y.~-S.~Liu,
  arXiv:1012.0453 [nucl-th].

\end{thebibliography}
\end{document}